\documentclass{webofc}
\usepackage[varg]{txfonts}   
\begin{document}
\title{Gaining insight from large data volumes with ease}
%
%

\author{\firstname{Valentin} \lastname{Kuznetsov}\inst{1}\fnsep\thanks{\email{vkuznet@gmail.com}}}

\institute{Cornell University, Ithaca, NY, USA 14850}

\abstract{%

Efficient handling of large data-volumes becomes a necessity in today's world. 
It is driven by the desire to get more insight from the data and to gain a better
understanding of user trends which can be transformed into economic incentives
(profits, cost-reduction, various optimization of data workflows, and
pipelines). In this paper, we discuss how modern technologies are transforming
well established patterns in HEP communities. The new data insight can be
achieved by embracing Big Data tools for a variety of use-cases, from
analytics and monitoring to training Machine Learning models on a terabyte
scale. We provide concrete examples within the context of the CMS experiment where
Big Data tools are already playing or would play a significant role in daily operations.
}
\maketitle

\section{Introduction}

With the CERN LHC program underway, we start seeing an exponential acceleration
of data growths in the High-Energy Physics (HEP) field\footnote{Here we refer to the data as
raw and MonteCarlo (MC) data produced and processed by experiments as well as
the associated meta-data and distinguish them explicitly in the paper.}.
In Run II CERN experiments operated in the
petabyte (PB) regime. For instance, in 2017 the CMS experiment alone produced around 30
billion raw
events and complemented them with 16 billion Monte Carlo similated events. It successfully transferred
a few PB/week with average transfer rates of 2-6 GB/s; almost 20PB of data were
replicated at GRID T1 sites and about 80PB at T2 sites.  The disk utilization
was at the level of 20/40/50PB at T0/T1/T2 sites respectively. With the up-coming
High Luminosity LHC (HL-LHC) program at CERN all HEP experiments will face a new
challenge, the exabyte ($10^{18}$) era of computing \cite{HEP-CWP}. We
anticipate that new techniques and technologies will be required to handle this
unprecedented amount of data. For example, the overall time of a typical physics
analysis may be significantly reduced by moving away from sequential processing
of events at GRID data-centers to data-reduction facilities based on Big Data
technologies \cite{HEP-BigData}. Similarly, the experiment's meta-data and services
are undergoing significant changes by embracing data-processing on Hadoop+Spark
platforms, and adding NoSQL databases with their traditional RDBMS counterparts
to a growing list of data-services.

In this paper, we discuss new technologies and techniques for handling
experiment a meta-data to gain additional insight from distributed data sources,
located at data-centers, on HDFS, in relational and NoSQL databases.  The new
information obtained with parallel processing of large datasets helps us
better understand our resources, more efficiently utilize computing
infrastructure, and gradually move towards a data-driven approach in the HL-LHC era of
computing.

\section{Current landscape}

When the concept of relational databases was introduced in 1970 \cite{RDBMS} it
solved many problems in the world of information technology. In the HEP community the
RDBMS technologies were used in many data services from online calibration
to offline data bookkeeping systems as well as surrounding
infrastructure.  Most HEP experiments successfully adopted RDBMS technologies
(open-source and commercial products) for their needs. At CERN, almost all
production systems rely on ORACLE databases. For instance, in CMS we use it
for dozens of data-services, and the two largest databases, DBS and PhEDEx
\cite{CMSDataManagement}, are around a few hundred GBs each excluding indexes.

During Run II operations we start seeing a limitations of RDBMS based solutions
to address the experiment needs. For instance, in CMS, users are required to place
queries across multiple databases to find their desired information.  To
overcome these obstacles, the CMS experiment has developed a Data Aggregation
System \cite{CMS-DAS}. It was designed as an additional layer above existing
data-services (based on RDBMS backends) and used an NoSQL (MongoDB \cite{MongoDB})
database as a caching layer. It aggregates information from different
data-services, and presents it to end-users via the flexible Query Language based on
the SQL syntax without explicitly requiring joints among database tables. Even though it was
successfully used by CMS in production for many years the
growth of information in the experiment requires new solutions to address
increasing demand for information, e.g.  for monitoring purposes.  By the end of
Run II, users started becoming more interested in a new type of aggregated information which
requires data-processing among distributed databases, joining various
attributes, and spans across large datasets. A typical example would be data
popularity plots of user activities for all T1 and T2 sites over large period
of time, e.g. up to a year.  To extract this information, it was required to
join almost all tables among the three largest databases: the CMS Data Bookkeeping
System (DBS), which tracks all experimental datasets, the PhEDEx database, which
knows about data location, and the data popularity database which keeps track of user
jobs run at various data-centers.  Such tasks cannot be accomplished via SQL queries
since the information physically resides in different databases, and even though
ORACLE tools provide the ability to perform cross-database joins we found that it
does not scale well in practice.  Therefore, the process requires manual
extraction of relevant information from all databases, proper
data-preprocessing, and a complex workflow to obtain desired results. Later, we
realized that such a workflow can be easily resolved if all required data will
reside on HDFS where we can apply the Spark framework \cite{Spark} over
distributed dataframes and take advantage of parallel processing on a HDFS
cluster. Data placement of various meta-data sources started in early 2015
and includes several CMS databases, HTCondor logs, CMSSW file access logs, file
transfer records as well as Workflow Management logs.  At the moment, the CMS
experiment has migrated dozens of data sources to HDFS (Table.
\ref{table:CMS-HDFS-Data}) and accumulated more than 32 TB of data
stored in various data-formats.

\begin{center}
\begin{table}[htb]
\centering
\begin{tabular}{lll}
    \hline\hline
    HTCondor logs \cite{CMS-HTCondor} & JSON & 11.1 TB \\
    AAA (Global Data Access) logs \cite{CMS-AAA} & JSON & 11 TB \\
    EOS logs \cite{EOS} & JSON & 5.3 TB \\
    FTS (File Transfer System) logs \cite{FTS} & JSON & 4.2 TB \\
    PhEDEx snapshots \cite{CMSDataManagement} & CSV & 3.3 TB \\
    WMArchive logs \cite{CMS-WMArchive} & Avro & 1.3 TB \\
    CMSSW (CMS SoftWare framework) logs & Avro & 0.5 TB \\
    DBS tables \cite{CMSDataManagement} & CSV & 0.3 TB \\
    JobMonitoring logs & Avro & 0.2 TB \\
    \hline\hline
\end{tabular}
    \caption{\label{table:CMS-HDFS-Data}Current snapshot of CMS meta-data on HDFS stored on HDFS.}
\end{table}
\end{center}

The data on HDFS are stored in various data-formats which are suited for
different purposes, e.g. log files are usually streamed to HDFS in native
data-format (JSON), the database tables are easily converted into CSV
data-format, while large unstructured data sets, e.g. in case of the WMArchive system
\cite{CMS-WMArchive}, are converted into compact, fast,
binary Avro data-format with a pre-defined schema. Fortunately, the HDFS libraries
support a broad variety of data-formats, and the Spark framework is guaranteed to
work seamlessly and efficiently with all of them.

Such availability of large datasets and efficient processing on Hadoop clusters
open up new possibilities to push the boundaries of analytics tasks beyond
traditional approaches based on relational databases. The run time to spawn
TB of data using the Spark framework on HDFS is of the order of a couple of minutes
and is not restricted to the content of a single database. Multiple sources
can be combined and efficiently processed.

\section{New approaches}
New approaches to handle large datasets in the HEP community are emerging from
the business world. First, the NoSQL solutions are adopted to
allow storage of unstructured documents, support distributed natures of actors,
and information replication.  For example, in CMS we successfully adopted
MongoDB \cite{MongoDB} and CouchDB \cite{CouchDB} technologies for different
use-cases. The former is used as caching and persistent layers in the Data
Aggregation system \cite{CMS-DAS}, while the latter is successfully used in Data Management
and Workflow Management system \cite{WMAgent} to continuously replicate
workflows across distributed agents handling CMS MonteCarlo production.

\begin{figure}
\centering
\includegraphics[width=0.8\textwidth]{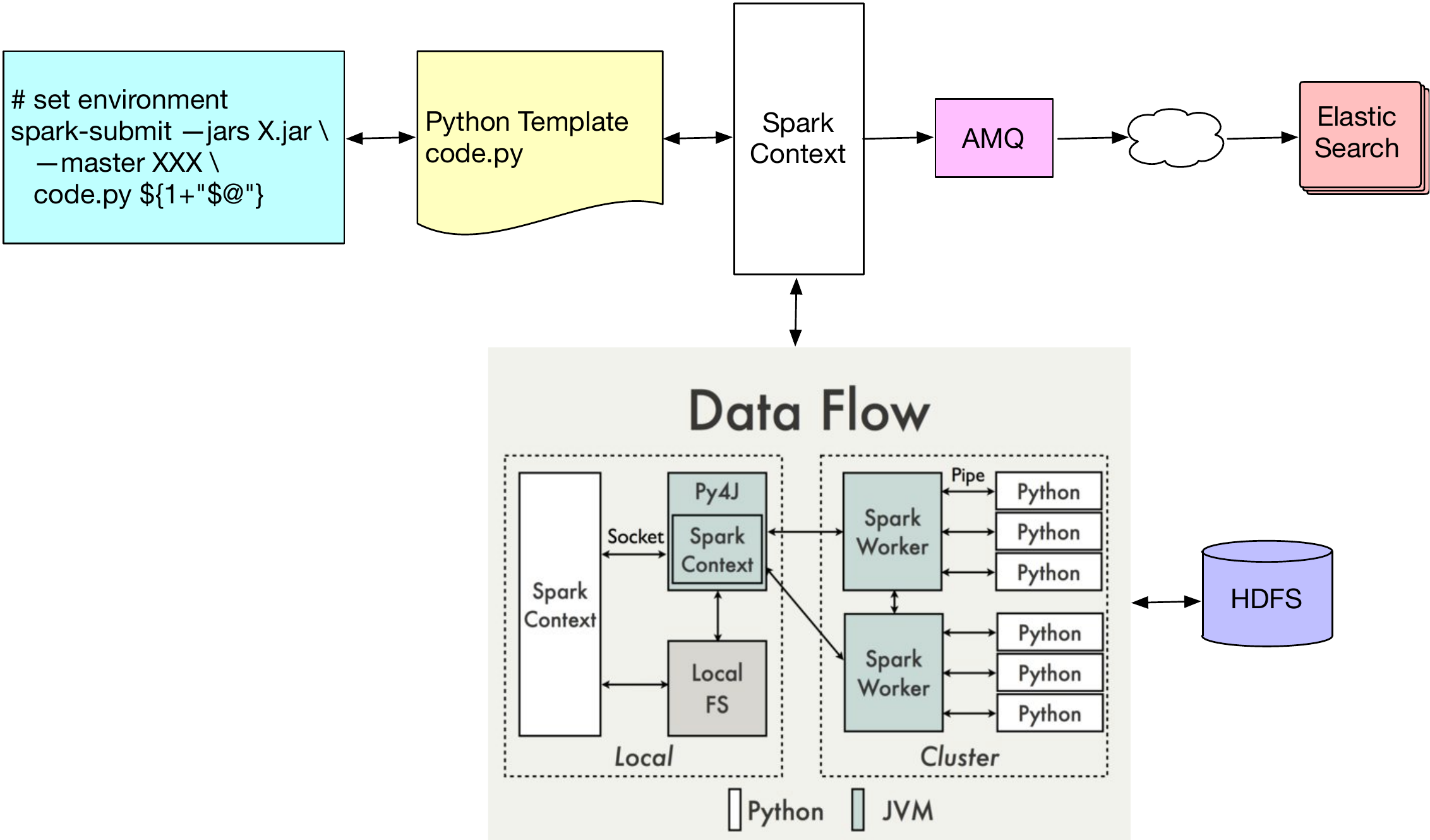}
    \caption{A typical CMSSpark workflow to process data on HDFS cluster via the
    Spark framework.}
\label{fig:CMSSparkWorkflow}
\end{figure}

As we mentioned in the previous section the Hadoop eco-system starts playing a
significant role in almost every HEP experiment.  Moreover, the CERN central
monitoring system (MONIT) \cite{CERNMonit} heavily relies on it and incorporates
various technologies and tools in their stack, e.g. Kafka, EalsticSearch, InfluxDB, Kibana, Grafana,
etc. But all of these innovations come with their own price.
Users are required to learn a broad variety of new tools, data-formats, etc., and
understand how to run their workflows in such an environment.  And, experiments
need to adopt their tools and data management systems to new technologies, too.
For example, in CMS a typical Spark workflow is quite complex task, see
Fig. \ref{fig:CMSSparkWorkflow}.
We rely on Python Spark APIs, to read and pre-process data from multiple
data-providers on HDFS. The aggregated information is often placed
back into the Asynchronous Message Queuing system and ends-up either in
ElasticSearch engine or CERN MONIT systems. Obviously, such a workflow
is difficult to construct properly and even harder to maintain on the long
run. Often, the majority of tasks was repeated among different users,
and a new level of abstraction was desired.

We simplified user access to HDFS, Spark and the CERN Hadoop eco-system via
the CMSSpark framework \cite{CMSSpark} which takes care of setting up a cluster
environment, provides a layer of abstraction to data access on HDFS, performs
data-format transformation to Spark
DataFrames, handles job handling and data placement back to HDFS and/or CERN MONIT
systems. At the end, users are required only to write an analysis code to process
the desired DataFrames. A submission of user tasks to the CERN HDFS cluster
is simplified to the following command:

\begin{verbatim}
# shell_wrapper + user_workflow + user options
run_spark workflow.py --date 20180812 --fout hdfs:///cms/users
\end{verbatim}

Such simplicity has boosted the adaptation of Hadoop tools within the CMS
collaboration and was quickly adapted to a variety of use-cases, see the discussion
in Sect. \ref{CMS-monitoring}.

Although the discussion above was applied to meta-data, the Big Data tools can also help
the HEP community apply new techniques to process and analyze real data. For
instance, the authors of \cite{HEP-BigData} proposed to use HDFS/Hadoop as a
data-reduction facility in HEP analysis with the ambitious goal of reducing 1PB of
raw input data to 1TB output data in a few hours. This approach can be further
extended not only to HEP analysis per-se, but also to train Machine Learning
(ML) models on petabyte datasets.

\begin{figure}
\centering
\includegraphics[width=0.8\textwidth]{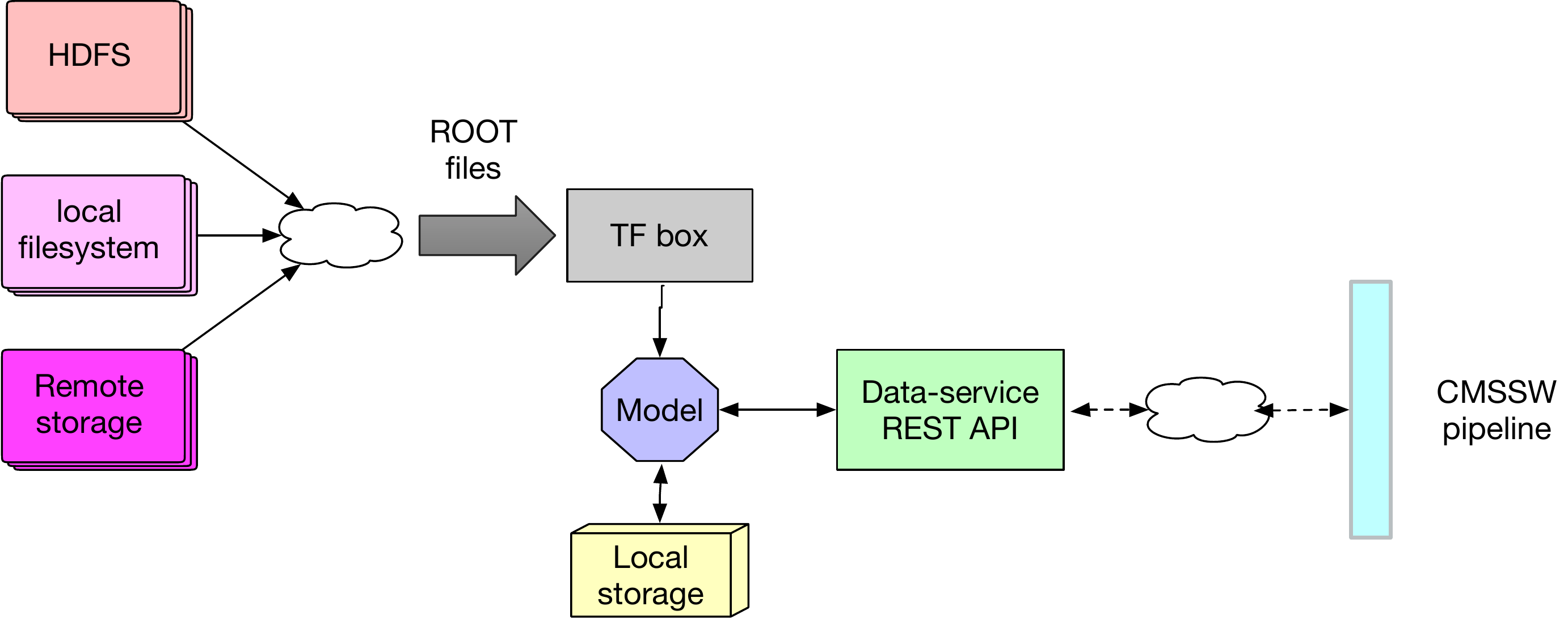}
    \caption{A TensorFlow as a Service \cite{TFaaS} architecture for a HEP use case.
    The input raw data in ROOT data-format can be read from remote
    data-providers, including XRootD servers, local filesystems, and HDFS,
    be processed and transformed on the HDFS/Spark platform and fed
    into ML framework for training (e.g. via SparkML). The trained model
    can be served to end-users or the entire framework (CMSSW) as an HTTP based
    data-service.}
\label{fig:HEP-BigData-ML}
\end{figure}

The ML models in the HEP community have been used for years, e.g. Boosted Decision
Trees or simple Neural Networks which are often used in various physics analysis. The
recent advances of technologies both on hardware and software fronts allow ML
models to be adopted universally, from computing infrastructure to trigger systems
\cite{HEP-ML-CWP}.  But the key problem with ML training is the data preparation
step which involves data transformation and pre-processing. The most common
data-format for ML training is CSV (or alike) while the HEP data are stored in ROOT
data-format.  Recent developments in ROOT I/O \cite{ROOT-IO} and ROOT data
access \cite{DataAccess} open up a possibility to directly read and process
ROOT data on the HDFS cluster. With this change we are already able to organize new
types of workflows of reading petabytes of
data, perform necessary data-transformation and pre-processing on HDFS,
train ML models, and deliver them to end-users as a data-service \cite{TFaaS}.
Fig. \ref{fig:HEP-BigData-ML} demonstrates such an R\&D pipeline in the context
of the TFaaS project for the CMS experiment.

Preliminary studies shows that we can achieve reading HEP events at the rate of
$~100$kHz (50 MB/s) via the uproot \cite{uproot} library, pre-process TB of data
in a range of minutes to a few minutes to few hours\footnote{The processing time
strongly depends on the specific use-case and complexity of the executed workflow.}
on the HDFS cluster \cite{CMSMLSpark}, train a
model and serve it to end-users via TensorFlow as a Service \cite{TFaaS} tool
with high throughput in a distributed environment. Our benchmarks have shown
that we can easily achieve $~500$ req/s throughput for concurrent clients
using a single node serving TFaaS data-service.

These new approaches are developed rapidly and are partially adopted
in the CMS experiment. Below we briefly discuss a few of them in the context
of the CMS monitoring infrastructure.

\subsection{CMS monitoring\label{CMS-monitoring}}
The CMS monitoring infrastructure is gradually migrating from experiment specific tools to
the central CERN Monitoring system \cite{CERNMonit}. The CERN MONIT system consists
of more than a hundred data producers with 3.5 TB/day injection rate
and dozens of Spark jobs running 24/7. It is highly
integrated with the CERN Analytix cluster composed of 39 nodes with 64GB of RAM, 32
cores/node Mix of Intel® Xeon® CPU E5-2650 @ 2GHz AMD Opteron(™) 6276, and
capable of storing and handling PB of data.

So far we have migrated several experiment dashboards to the CERN MONIT
infrastructure, among them AAA, EOS, HTCondor, task monitoring of user
analysis jobs as well as WMArchive sub-systems. We found that the Spark platform
significantly improved our analytics capabilities.
For instance, in the WMArchive \cite{CMS-WMArchive} system we can promptly perform
the following tasks:
\begin{itemize}
    \item identify failed workflows and problematic sites
    \item spot production issues via log look-up and exit codes
    \item monitor CMS production status, including sites, campaigns monitoring and extracting throughput metrics
    \item perform data aggregation and produce aggregated statistics.
\end{itemize}
The system was designed to collect 100M+ documents per year from distributed
WMAgents with an upload rate of $O(1M)$ documents per day. The documents were injected into
the local cache of MongoDB and are transferred to HDFS for long-term storage.
We periodically run daily and hourly aggregation jobs to gain insight
on MonteCarlo production workflows. This information is fed into
the CERN MONIT system where it is displayed in various dashboards.

Using CMS data on HDFS, as outlined Table \ref{table:CMS-HDFS-Data},
we analyzed the most popular data tier among end-users in
2017. To our satisfaction it was MINIAODSIM accessed by 56\%, 42\%, 40\%, 40\%
in AAA, EOS, CMSSW, and CRAB systems, respectively. Then it was followed
by MINIAOD, AOD, and RAW data-tiers. The usage of the RECO data-tier was quite
negligible, at the level of a few percent in corresponding systems.


We also look at the data popularity content of our data and successfully used
the Hadoop Spark platform as a data reduction and processing facility.
We performed studies to predict
dataset popularity using user AAA logs \cite{CMSMLSpark}. We demonstrated that
it can be modeled via ML and be used as a seed by the CMS dynamic
data placement system. It worth mentioning that the original dataset had 2B rows of
AAA logs combined with PhEDEx database tables.  This dataset was reduced to
0.5M records in a couple of hours and fed into the SparkML framework for training the ML
model.  We foresee that such a workflow pipeline can be successfully adopted in
the HL-LHC era where intelligent data placement may play a critical role.

Finally, we used Job Monitoring and WMArchive data to measure sites
performance. The studies targeted a concrete architecture on T2 sites
where node throughput was calculated as a number of processed events per
second for various processor architectures taking into account the number of job
slots per core. These results complemented well established HS06 scores and
were included in the HEPiX Benchmarking Working Group \cite{HEPiX}.

\section{Summary}
The CMS experiment is continuously improving its computing and offline
infrastructure. In particular, it is shifting its monitoring infrastructure
to the central CERN monitoring system and has a large number of ML projects.

In this paper we discussed a gradual shift to handle large datasets in the CMS
experiment. The large portion of CMS meta-data has been successfully migrated
to HDFS and complements our existing database solutions. The usage of Hadoop tools
and its eco-system is no longer a problem due to the simplicity of the CMSSpark
framework. It simplified access to a broad variety of CMS data located on HDFS
by abstracting the data access layer, and provided a simple and uniform
way to submit, process, and analyze these data.

We also described a new use-case for Machine Learning training over large
distributed datasets and are continuously delivering ML models to end-users via a new
Tensor as a Service data platform recently developed and which is currently undergoing testing in
the CMS experiment. Such a service may not only serve the CMS
experiment per-se but can be applicable to other experiments. It may
fill the gap of integrating ML tools into existing infrastructure and
experiment framework. Finally, we foresee that this approach will gain 
popularity in the upcoming years of data taking in the HL-LHC regime where training ML
models over petabyte datasets will become a norm.

\begin{acknowledgement}
I would like to thank my CMS colleagues David Lagne (Princeton)
and Carl Vuosalo (Univ. of Wisconsin) for their support and
various contributions in CMS monitoring infrastructure, including
production and validation of of T1/T2 usage plots.
I also would like to thank Luca Menichetti from CERN IT who provided support
for development, maintenance and deployment of our scripts on the Spark platform.
Special thanks go to the CERN MONIT team for collaboration
and support to set up and maintain the CMS Monitoring dashboards.
\end{acknowledgement}

%

\begin{thebibliography}{}
%

\bibitem{HEP-CWP}
    A. A. Alves Jr., et. al.,
        {\it A Roadmap for HEP Software and Computing R\&D for the 2020s},
    https://arxiv.org/abs/1712.06982

\bibitem{HEP-BigData}
    O. Gutche, et al.,
        {\it Big Data in HEP: A comprehensive use case study},
    doi: 10.1088/1742-6596/898/7/072012, 
    https://arxiv.org/abs/1703.04171

\bibitem{RDBMS}
    E.F. Codd, {\it A Relational Model of Data for Large Shared Data Banks},
    Communications of the ACM. 13 (6): 377–387. doi:10.1145/362384.362685.

\bibitem{CMSDataManagement}
	M. Giffels, Y. Guo, V. Kuznetsov, N. Magini and T. Wildish
	{\it The CMS Data Management System}
	J. Phys.: Conf. Ser. 513 (2014) 042052; doi:10.1088/1742-6596/513/4/042052

\bibitem{CMS-DAS}
    V. Kuznetsov, D. Evans, S. Metson,
    {\it The CMS Data Aggregation System},
    doi:10.1016/j.procs.2010.04.172

\bibitem{MongoDB} MongoDB document-oriented database,
    https://docs.mongodb.org/

\bibitem{Spark} Apache Spark, https://spark.apache.org/

\bibitem{CMS-HTCondor}
	D. Thain, T. Tannenbaum and M. Livny (2005),
	{\it Distributed computing in practice: the Condor
			experience Concurrency and Computation: Practice and Experience}
	17 2-4 323-356 doi:10.1002/cpe.938

\bibitem{CMS-AAA}
    K. Bloom, et. al.,
    {\it Any Data, Any Time, Anywhere: Global Data Access for Science}, BDC 2015: 85-91
    https://arxiv.org/pdf/1508.01443.pdf

\bibitem{EOS}
    X. Espinal, et. al.,
    {\it Disk storage at CERN: Handling LHC data and beyond}
    Journal of Physics: Conference Series 513 (2014) 042017 doi:10.1088/1742-6596/513/4/042017

\bibitem{FTS}
    A A Ayllon, et. al.,
        {\it FTS3: New Data Movement Service For WLCG},
        2014 J. Phys. Conf. Ser. 513 032081

\bibitem{CMS-WMArchive}
    V. Kuznetsov, N. Fischer, Y. Guo,
        {\it The archive solution for distributed workflow management agents of the CMS experiment at LHC}
        Computing and Software for Big Science 2018, 2:1, doi: 10.1007/s41781-018-0005-0

\bibitem{CouchDB} Apache CouchDB data-management system,
    http://couchdb.apache.org/

\bibitem{WMAgent}
	D. Evans, et. al., {\it The CMS workload management system},
	Journal of Physics: Conference Series, Volume 396, Part 3

\bibitem{CERNMonit}
    A. Aimar, et. al.,
        {\it Unified Monitoring Architecture for IT and Grid Services}
        Journal of Physics Conference Series 2017, 898 092033, doi: 10.1088/1742-6596/898/9/092033

\bibitem{CMSSpark}
    V. Kuznetsov, {\it CMSSpark a general purpose framework to run CMS experiment workflows on HDFS/Spark platform}
    DOI 10.5281/zenodo.1401228
        https://zenodo.org/badge/latestdoi/74044584

\bibitem{HEP-ML-CWP}
    K. Albertsson, et. al.,
    {\it Machine Learning in High Energy Physics Community White Paper}
    https://arxiv.org/abs/1807.02876

\bibitem{ROOT-IO}
    B. Bockelman, Z. Zhang, J. Pivarski,
    {\it Optimizing ROOT IO For Analysis},
    https://arxiv.org/abs/1711.02659

\bibitem{DataAccess}
    J. Pivarski, P. Elmer, B. Bockelman, Z. Zhang,
    {\it Fast Access to Columnar, Hierarchical Data via Code Transformation}
    CoRR abs/1708.08319 (2017)

\bibitem{TFaaS}
    V. Kuznetsov, {\it TensorFlow as a Service} doi: 10.5281/zenodo.1308048.

\bibitem{uproot}
    DIANA-HEP Scikit-hep uproot library, {\it Minimalist ROOT I/O in pure Python and Numpy},
    https://github.com/scikit-hep/uproot

\bibitem{CMSMLSpark}
    M. Meoni, V. Kuznetsov, L. Menichetti, J. Rumševičius, T. Boccali, D. Bonacorsi,
    {\it Exploiting Apache Spark platform for CMS computing analytics},
    ACAT 2017, http://arxiv.org/abs/1711.00552

\bibitem{HEPiX}
    HEPiX Benchmark and Performance group,
    https://w3.hepix.org/benchmarking.html

\end{thebibliography}
%
%

\end{document}